# Excitonic Effects on the Optical Response of Graphene and Bilayer Graphene

Li Yang, Jack Deslippe, Cheol-Hwan Park, Marvin L. Cohen, and Steven G. Louie

Department of Physics, University of California at Berkeley, California 94720 and Materials Sciences Division, Lawrence Berkeley National Laboratory, Berkeley, California 94720

Abstract:

We present first-principles calculations of many-electron effects on the optical response of graphene, bilayer graphene, and graphite employing the GW-Bethe Salpeter equation approach. We find that resonant excitons are formed in these two-dimensional semimetals. The resonant excitons give rise to a prominent peak in the absorption spectrum near 4.5 eV with a different lineshape and significantly red-shifted peak position from those of an absorption peak arising from inter-band transitions in an independent quasiparticle picture. In the infrared regime, our calculated optical absorbance per graphene layer is approximately a constant, 2.4%, in agreement with recent experiments; additional low frequency features are found for bilayer graphene because of band structure effects.

Excitonic effects are observable in the optical response of semiconductors. Earlier work based on a tight-binding bond-orbital model illustrated these effects in bulk semiconductors [1] and *ab initio* methods employing the GW-Bethe Salpeter equation (GW-BSE) approach have become available in the past decade [2]. Excitonic effects are commonly believed to be unimportant in the optical spectrum of metals because of strong screening. However, recent first-principles calculations [3, 4] have predicted, and subsequent experimental studies [5] have confirmed, the existence of bound excitons in one-dimensional (1D) metallic carbon nanotubes (CNTs). Therefore, it is of considerable interest to explore whether there are significant excitonic effects in 2D metallic or semi-metallic systems.

Graphene is a 2D semimetal with interesting physics associated with its unusual electronic structure and its promising device applications [6-9]. In particular, the optical properties of graphene display many intriguing features, such as a near constant optical conductivity in the infrared regime and gate-dependent optical absorbance [10-13]. However, there have been no first-principles studies to date of the optical properties of graphene including excitonic effects that are known to be important in reduced dimensional materials.

In this work, we have carried out first-principles calculations using a many-body Green's function theory to study the optical spectra of graphene and bilayer graphene. Following the approach of Rohlfing and Louie [14], we calculate the optical response of isolated single- and bi-layer intrinsic graphene in three stages: (i) we obtain the electronic ground state using density functional theory (DFT) within the local density approximation (LDA); (ii) the quasiparticle excitations are calculated within the GW approximation [15]; and (iii) we solve the Bethe-Salpeter equation (BSE) to obtain the photo-excited states and optical absorption spectrum [1, 2, 14].

Our first-principles results on graphene show that, for single-particle excitations, there is a significant self-energy correction to the band velocity of the Dirac quasiparticles. Owing to electron-hole interactions, the absorption peak arising from the interband transitions at around 5.1 eV is totally suppressed and replaced by a new peak at 4.5 eV with a very different lineshape. This change in the optical spectrum is the result of a redistribution of optical transition strengths by strong resonant excitons. These results persist in bilayer graphene. Moreover, the calculated infrared spectral absorbance per graphene layer including electron-hole interactions is approximately a constant, 2.4%, in agreement with recent experiments [10, 11].

In our studies, the intra-layer structure of graphene and bilayer graphene is fully relaxed using the calculated forces and stress on the atoms within DFT/LDA. For Bernal-stacked bilayer graphene, the inter-layer distance is chosen to be the experimental value of graphite (0.334 nm). The calculations are done in a supercell arrangement [16] with a plane-wave basis using norm-conserving pseudopotentials [17] with a 60 Ry energy cutoff. The distance between graphene sheets in neighboring supercells is 1.2 nm to avoid spurious interaction. A 32x32x1 uniform k-point grid is used to ensure converged LDA results and a 64x64x1 k-point grid is necessary for computing the converged self energy. We take into account dynamical screening effects in the self energy through the generalized plasmon pole model [15]. In solving the BSE, we made used two usual approximations: 1) the Tamm-Dancoff approximation, which has given accurate results for the optical absorption spectra of other metallic systems such as graphite and metallic CNTs [3-4], and 2) the static electron-hole interaction approximation since the excitation energy of the resonant excitons is large (~5 eV) relative to the electron-hole interaction energy. The electron-hole interaction kernel is evaluated first on a coarse k-grid (64x64x1) and then interpolated onto a fine grid (200x200x1) [14]. Two valence bands and two conduction bands are included for the optical absorption spectra up to 7.0 eV. Inclusion of more bands does not change the spectrum in the 0-7 eV range. In this Letter, we shall focus on the absorption spectrum for light polarized parallel to the graphene plane.

The LDA eigenvalues and GW quasiparticle energies of graphene close to the Dirac point are shown in Fig. 1 (a). While the LDA Fermi velocity of graphene is $0.85 \times 10^6$ m/s, the GW value is $1.15 \times 10^6$ m/s that is in good agreement with experiment [8] as well as with previous GW calculations [18, 19]. The LDA and quasiparticle band structures of bilayer graphene are shown in Fig. 1 (b).

Figure 2(a) shows the calculated optical spectrum of graphene with and without electron-hole interaction included. The plotted quantity $\alpha_2(\omega)$ is the imaginary part of the polarizability per unit area and is obtained by multiplying the calculated dielectric susceptibility, $\chi = (\varepsilon-1)/4\pi$, by the distance between adjacent graphene layers (or bilayers) in our supercell arrangement. This quantity $\alpha_2(\omega)$ when multiplied by the area of graphene or bilayer graphene gives the polarizability of the system. The absorption below 0.3 eV is not shown because intraband transitions and temperature effects are important there and our calculation does not include these factors. In absence of electron-hole interaction, the interband transitions form a prominent absorption peak at 5.15 eV. However, with excitonic effects included, a prominent absorption peak now appears at 4.55 eV which is a 600 meV apparent shift. In addition, the peak profile in both $\alpha_2(\omega)$ and the absorbance is substantially modified from an almost symmetric peak in the interband transitions case to an asymmetric one in the exitonic case.

It is surprising to find such large excitonic effects (an apparent shift of 600 meV and a dramatic change in shape of the optical peak) in this 2D semimetal, given that the binding energy of excitons found in 1D metallic carbon nanotubes is only tens of meV and that there is no significant excitonic effect in bulk metals. In Fig. 2 (b), we see that both the joint density states (JDOS) of quasiparticles from the GW calculation and the density of excitonic states from solving the BSE are nearly identical, similar to findings in bulk semiconductors [14]. These changes arise mainly from the attractive direct term in the electron-hole kernel, with the repulsive exchange term plays a negligible role.

To analyze our results, we rewrite the relevant optical transition matrix element for going from the ground state $|0\rangle$ to a correlated electron-hole (exciton) state $|i\rangle = \sum_k \sum_v^{hole} \sum_c^{elec} A^i_{vck} |vck\rangle$ into the form

$$\langle 0|\vec{v}|i\rangle = \sum_v \sum_c \sum_k A^i_{vck} \langle vk|\vec{v}|ck\rangle = \int S_i(\omega) d\omega, \quad (1)$$

where

$$S_i(\omega) = \sum_v \sum_c \sum_k A^i_{vck} \langle vk|\vec{v}|ck\rangle \delta(\omega - (E_{ck} - E_{vk})), \quad (2)$$

which gives a measure of the contribution of all interband pairs (*ck*, *vk*) at a given transition energy ω to the optical strength of the exciton state *i*. Because of inversion symmetry of graphene, $S_i(\omega)$ is given as a real function. In Fig. 3, $S_i(\omega)$ and its integrated value up to a given frequency are depicted for three optically bright excited state to provide an understanding of why excitonic effects enhance the absorption around 4.5 eV but depress it around 5.1 eV.

Figure 3 (a) shows $S_i(\omega)$ for a state at 1.6 eV. Since there is negligible excitonic effect below 2.0 eV, this state displays a very narrow energy distribution in $S_i(\omega)$. In Fig. 3 (b) and (c), the states studied, located around 4.5 eV and 5.1 eV, respectively, show a considerably wider energy distribution in $S_i(\omega)$, indicating that they are linear combinations of many free electron-hole pair configurations of different energies which is consistent with having significant excitonic effects in this energy regime [20]. (The real-space wavefunctions of two resonant states are given as online supplementary figures to this article. See EPAPS.)

In Fig. 3 (b), the optical interband transition matrix element distribution for the state has larger amplitude extending to the high-energy direction. Since the running integrated value of $S_i(\omega)$ in Fig. 3 (b) increases significantly from 4.5 eV to 5.5 eV which is exactly the range of the strong interband absorption peak, this particular resonant exciton steals optical transition strength from the prominent interband absorption peak around 5.1 eV and enhances the optical absorption around 4.5 eV. In contrast, the $S_i(\omega)$ for the state shown in Fig. 3 (c) is more anti-symmetric than that shown in (b). The negative and positive contributions above and below 5.1 eV, respectively, nearly cancel each other and the final integrated strength is only one third of that for the exciton in (b). As a result, the optical absorption around 5.1 eV is depressed. Thus, similar to 1D graphene nanostructures [3-5, 21], excitonic effects dominate the optical absorption spectra in the energy region from 4 to 5 eV in 2D graphene itself.

The absorbance of graphene is expected to be a constant (2.29%) in the infrared spectral range [10-12, 22-28]. Figure 2(c) shows our calculated absorbance, $A(\omega) = \frac{4\pi\omega}{c}\alpha_2(\omega)$, of graphene with and without excitonic effects included. The absorbance is nearly the same (around 2.4%) in both cases. Our result is in good agreement with measured values and is consistent with previous studies [29]. Figure 2(d) compares the calculated results to experimental data [10, 11]. The calculated absorbance has a small but finite slope, in good agreement with measurements in Ref. 11. The very small excitonic influence in the infrared regime may be attributed to a vanishing joint density of states near zero energy.

The absorbance of bilayer graphene is presented in Fig. 4 (a), showing similar excitonic effects as those in single-layer graphene, with a significant red-shift in the position of the prominent absorption peak around 5 eV. There is also a noticeable absorption feature at 0.4 eV, contributed by interband transitions near the Dirac point. Moreover, the infrared spectral absorbance is around 4.8%, twice that of graphene.

The large excitonic effects in bilayer graphene near 4.5 eV have similar origins as those in graphene. However, as shown in Fig. 1 (b), the lowest conduction band and the highest valence band in the bilayer case touch each other and have parabolic shape. This provides a larger DOS around the Fermi level and results in stronger screening than in intrinsic graphene. Therefore, as shown in Fig. 4 (a), an apparent 450 meV red-shift of the prominent absorption peak of bilayer graphene is obtained with the inclusion of electron-hole interaction, which is smaller than that of graphene (600 meV).

We give also our calculated optical absorption of graphite together with experimental results [30] in Fig. 4 (b) and summarize the main absorption peak position of graphene, bilayer graphene, graphite and experimental data in Table I. Our results for graphite with excitonic effects are in good accord with experiment. We note that self-energy corrections and excitonic effects cancel each other more or less, making the positions of the main absorption peak of all three structures similar although the character of the excited states can change significantly. This near cancellation effect was noticed in other nanostructures previously [3, 31].

In conclusion, we have performed first-principles calculations on the quasiparticle energies and optical properties of single- and bi-layer graphene and graphite with many-electron effects included. A substantial self-energy renormalization of the Fermi velocity of graphene is obtained. Resonant excitonic effects in graphene and bilayer graphene result in significant changes in the optical absorbance spectrum in the energy regime near a van Hove singularity as compared to the independent-particle picture. Finally, we show that excitonic effects do not change the absorbance per graphene layer (for single- and bi-layer graphene) in the infrared range from that of the single-particle picture, in agreement with experimental findings.

We thank Y.-W. Son, D. Prendergast and E. Kioupakis for discussions. C.-H.P. and theoretical methods and codes were supported by NSF Grant #DMR07-05941, and L.Y. and J.D. and simulations studies were supported by the Director, Office of Science, Office of Basic Energy under Contract #DE-AC02-05CH11231. J.D. received a DOE

Computational Science Graduate Fellowship under Grant #DE-FG02-97ER25308. Computational resources provided by Lonestar of teragrid at the Texas Advanced Computing Center .

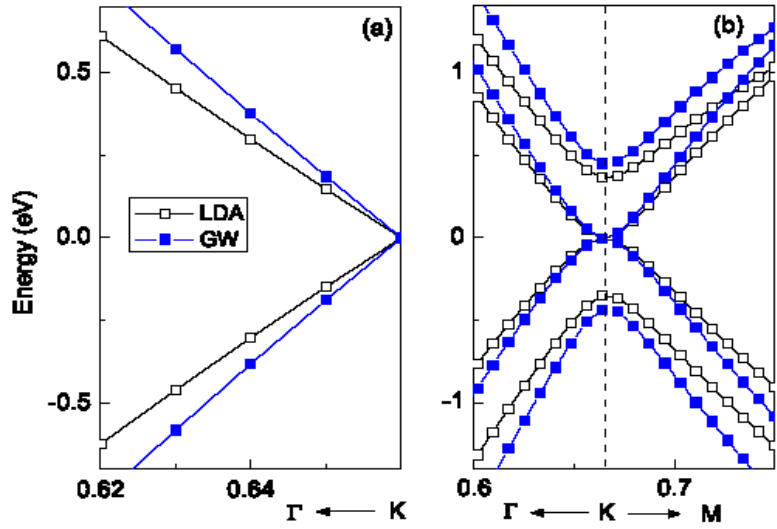

Fig. 1 (color online) LDA eigenvalues and GW quasiparticle energies of graphene (a) and bilayer graphene (b) close to the Dirac point. Wavevector is in $\frac{2\pi}{a}$, where $a$ is the in-plane lattice constant.

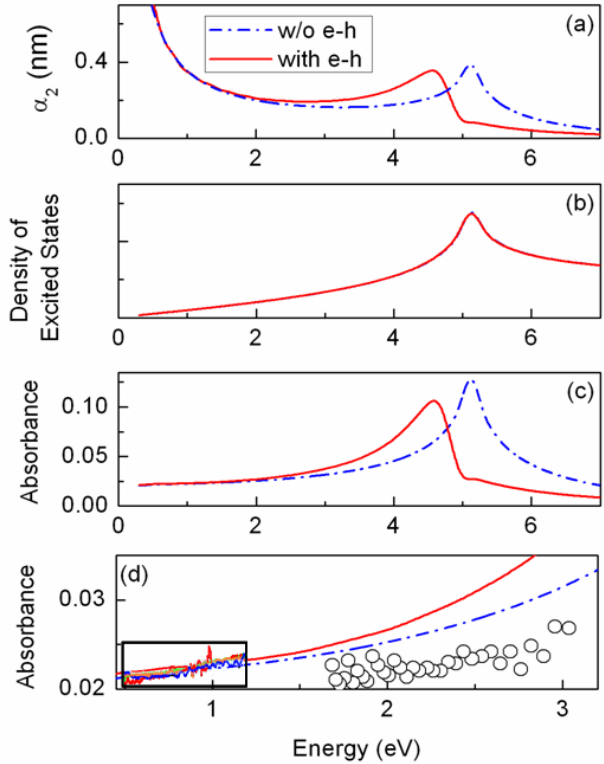

Fig. 2 (color online) (a) Optical absorption spectra, (b) density of excited states, (c) absorbance of graphene with (solid red curve) and without excitonic effects (dash-dot blue curve) included; and (d) comparison with experiments. In (d), rough colored curves within the small rectangular box are measurements from Ref. [11], and open circles are from those from Ref. [10]. A Lorentzian broadening of width of 0.05 eV is applied to all theoretical curves.

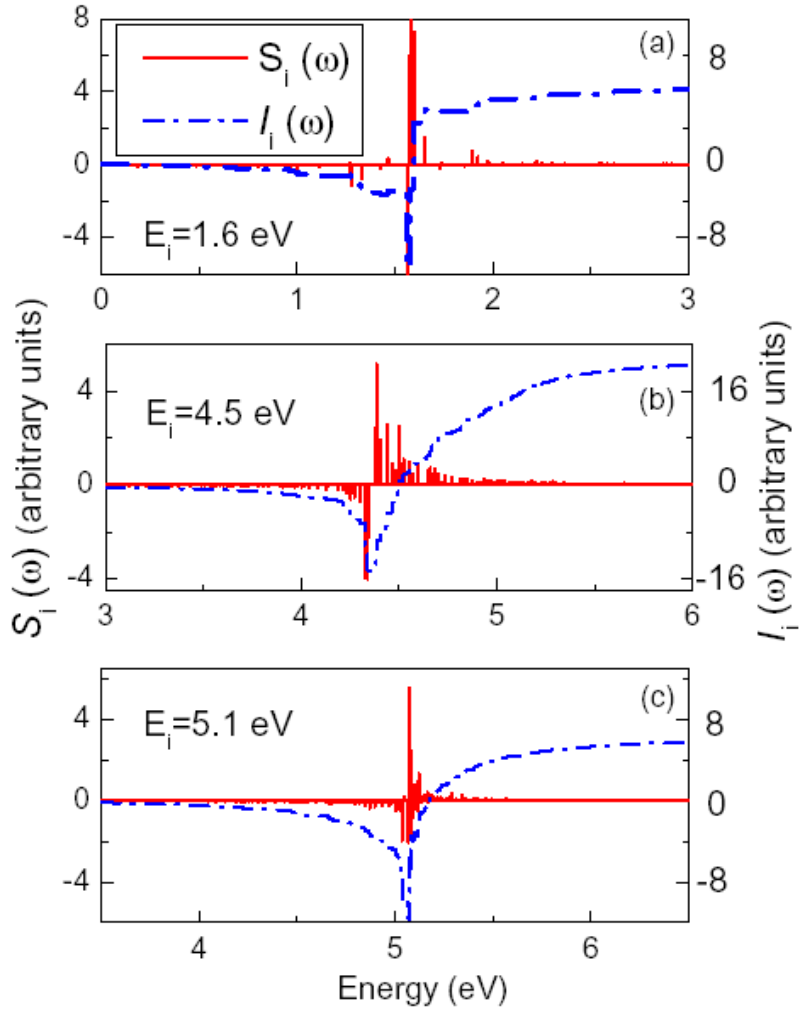

Fig. 3 (color online) $S_i(\omega)$ and the corresponding integral $I(\omega)=\int_0^\omega S(\omega')d\omega'$ of three optically bright states in graphene from GW-BSE calculations. The right scale of (b) is two times larger than that of (a) and (c).

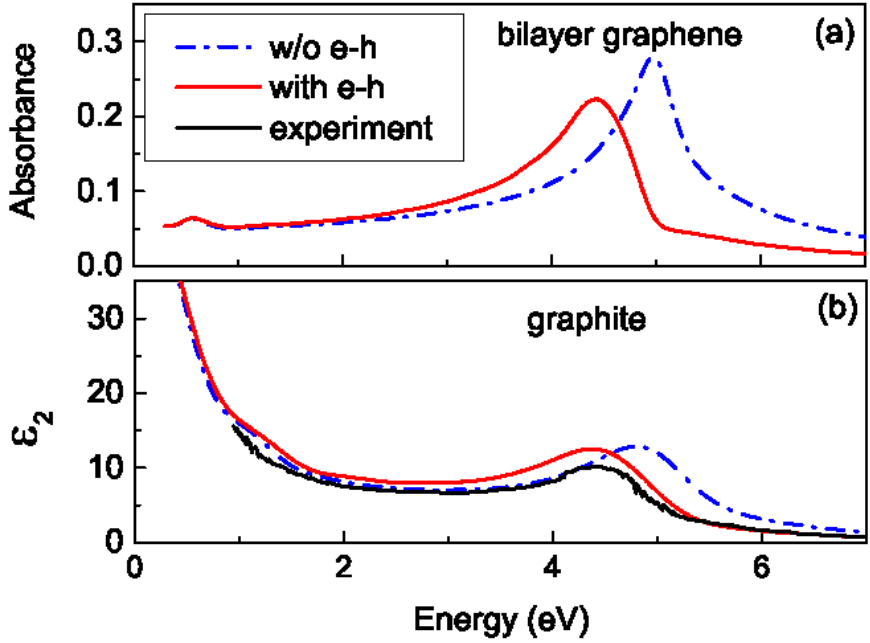

Fig. 4 (color online) (a) absorbance of bilayer graphene and (b) imaginary part of the dielectric function of graphite with and without excitonic effects included. Experiment data of graphite [30] are included in (b). A Lorentzian broadening with width of 0.05 eV has been applied to the theoretical curves.

TABLE I: Main absorption peak position (in eV) of graphene, bilayer graphene and graphite, and change in peak position rising from self-energy effects ($\delta\Sigma$) and from excitonic effects ($\delta_{exciton}$).

|  | graphene | bilayer graphene | graphite |
| --- | --- | --- | --- |
| $E_{peak}$ (Expt.) |  |  | 4.55 [30] |
| $E_{peak}$ (GW+BSE) | 4.55 | 4.52 | 4.50 |
| $\delta\Sigma$ | +1.10 | +0.91 | +0.69 |
| $\delta_{exciton}$ | −0.60 | −0.45 | −0.27 |